\documentclass[epj,amsfonts,amsmath,twocolumn,superscriptaddress,nofootinbib]{revtex4}

\usepackage{amssymb,color,paralist,amsmath,epsfig}
\usepackage{graphicx}
\usepackage{comment}
\usepackage{amsfonts}
\usepackage{relsize}
\usepackage{nicefrac,esint}
\usepackage{float}

\usepackage[colorlinks,citecolor=blue,linktoc=all,linkcolor=cyan]{hyperref}

\newcommand{\beq}{\begin{equation}}
\newcommand{\eeq}{\end{equation}}

\newcommand{\ber}{\begin{eqnarray}}
\newcommand{\eer}{\end{eqnarray}}

\begin{document}

\title{Two-photon exchange correction to the Lamb shift and hyperfine splitting of S levels}

\author{Oleksandr Tomalak}
\affiliation{Institut f\"ur Kernphysik and PRISMA Cluster of Excellence, Johannes Gutenberg Universit\"at, Mainz, Germany}
\affiliation{Department of Physics and Astronomy, University of Kentucky, Lexington, KY 40506, USA}
\affiliation{Fermilab, Batavia, IL 60510, USA}

\date{\today}

\begin{abstract}
We evaluate the two-photon exchange corrections to the Lamb shift and hyperfine splitting of S states in electronic hydrogen relying on modern experimental data and present the two-photon exchange on a neutron inside the electronic and muonic atoms. These results are relevant for the precise extraction of the isotope shift as well as in the analysis of the ground state hyperfine splitting in usual and muonic hydrogen.
\end{abstract}

\maketitle

\begin{figure}[h]
\begin{center}%\centering
\includegraphics[width=0.23\textwidth]{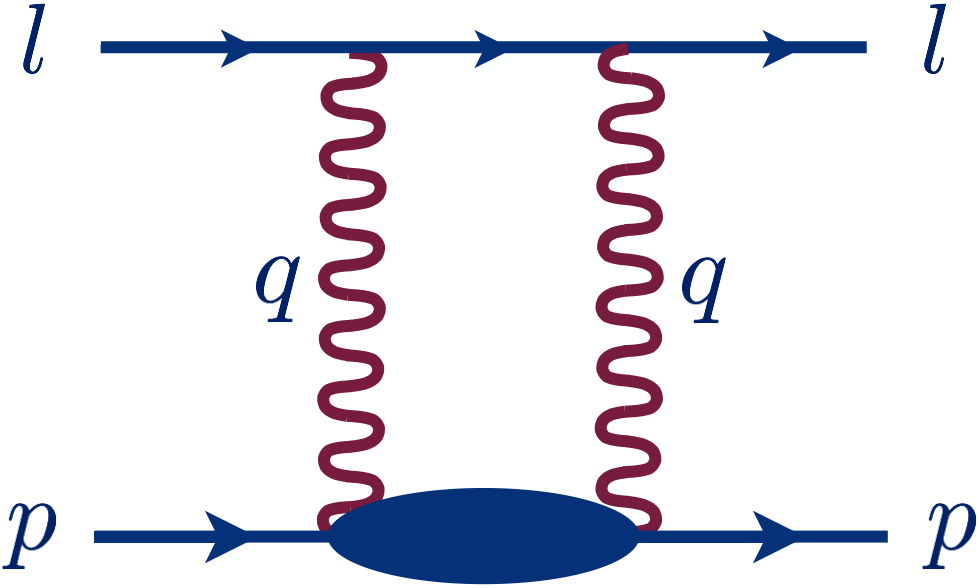}
\end{center}
\caption{Two-photon exchange graph.}
\label{TPE_graph}
\end{figure}

The discrepancy between the proton charge radius extractions from the Lamb shift in muonic hydrogen \cite{Pohl:2010zza,Antognini:1900ns} and electron-proton scattering \cite{Bernauer:2010wm,Bernauer:2013tpr,Mohr:2012tt} triggered a lot of theoretical and experimental efforts both in scattering and spectroscopy, see Refs. \cite{Carlson:2015jba,Hill:2017wzi} for recent reviews. Two-photon exchange (TPE) hadronic correction, see Fig. \ref{TPE_graph}, is a limiting factor extracting radii from the muonic hydrogen spectroscopy \cite{Pachucki:1996zza,Faustov:1999ga,Pineda:2002as,Pineda:2004mx,Nevado:2007dd,Carlson:2011zd,Hill:2012rh,Birse:2012eb,Miller:2012ne,Alarcon:2013cba,Gorchtein:2013yga,Peset:2014jxa,Tomalak:2015hva,Caprini:2016wvy,Hill:2016bjv,Carlson:2008ke,Carlson:2011af,Eides:2000xc}. Moreover, an accurate evaluation of two-photon corrections to hyperfine splitting (HFS) of ground state in electronic hydrogen in combination with an excellent experimental knowledge  \cite{Hellwig:1970,Zitzewitz:1970,Essen:1971x,Morris:1971,Essen:1973,Reinhard:1974,Vanier:1976,Petit:1980,Cheng:1980,Karshenboim:2005iy,Horbatsch:2016xx,Peset:2016wjq,Tomalak:2017lxo,Tomalak:2017npu,Tomalak:2017owk} (known with $\mathrm{mHz}$ accuracy) could help to analyse future precise measurements of 1S HFS in muonic hydrogen~\cite{Pohl:2016tqq,Dupays:2003zz,Adamczak:2016pdb,Ma:2016etb}, which aim to decrease an uncertainty of 1S-level HFS from the level of $40~\mu\mathrm{eV}$ \cite{Antognini:1900ns} up to the level of $0.2~\mu\mathrm{eV}$. Though the two-photon correction is smaller than the modern accuracy of Lamb shift measurements in usual hydrogen, it can affect the precisely measurable 1S-2S transition \cite{Parthey:2011yyy,Matveev:2013xxx} (with the experimental uncertainty 10-11$~\mathrm{Hz}$), as well as the isotope shift \cite{Parthey:2010aya,Jentschura:2011xx} (with the experimental uncertainty $15~\mathrm{Hz}$), above the accuracy level of the difference between proton and deuteron charge radii \cite{Eides:2000xc,Pachucki:2018yxe}. In the latter references, the elastic Friar term \cite{Friar:1978wv} was accounted for and the inelastic correction was estimated in the leading logarithmic approximation \cite{Bernabeu:1982qy,Khriplovich:1999ak,Friar:1997tr}. Besides two-photon corrections, the more involved three-photon exchange contribution to the Lamb shift was recently evaluated in the nonrecoil limit neglecting magnetic dipole and electric  quadrupole moments of the nucleus in Ref. \cite{Pachucki:2018yxe}.

In this paper, we provide a first complete dispersive calculation of $\alpha^5$ two-photon exchange contribution to the Lamb shift in electronic hydrogen, summarize the current status of this correction to the hyperfine splitting of S states and provide an update of Ref. \cite{Tomalak:2017lxo} for S-level HFS in $\mu\mathrm{H}$. Additionally, we present contributions to the Lamb shift arising from the two-photon exchange on the neutron inside a nucleus.

We evaluate the correction to the Lamb shift of S energy levels $\mathrm{E}_\mathrm{LS}$ following Refs. \cite{Pachucki:1996zza,Carlson:2011zd}. It can be expressed as a sum of three terms:
\ber
\mathrm{E}_\mathrm{LS} &=&  \mathrm{E}_{\mathrm{Born}} +  \mathrm{E}_{\mathrm{subt}} + \mathrm{E}_{\mathrm{inel}},
\eer
the Born contribution $\mathrm{E}_{\mathrm{Born}}$, the subtraction term $\mathrm{E}_{\mathrm{subt}}$ and the inelastic correction $\mathrm{E}_{\mathrm{inel}}$.
To evaluate the dimensionless forward unpolarized amplitude, we always normalize the TPE contributions to the energy $\mathrm{E}_0$:
\ber  \label{2S_correction_Lamb_shift}
\mathrm{E}_0 =  \frac{|\psi_{\mathrm{nS}}\left(0\right)|^2}{4 M m}, \label{normalization_energy}
\eer
where $M$ is the proton mass, $m$ is the lepton mass, $ | \psi_{\mathrm{nS}} (0) |^2 =  \alpha^3 m_r^3/(\pi \mathrm{n}^3) $ is the non-relativistic squared wave function of the hydrogen atom at origin with the reduced mass of the lepton and proton bound state $ m_r = M m / ( M + m )$, $\alpha$ is the fine-structure constant, and $\mathrm{n}$ is the principal quantum number. The inelastic contribution $ \mathrm{E}_{\mathrm{inel}}$ can be expressed as an integral over the unpolarized proton structure functions $F_1$ and $F_2$ \cite{Carlson:2011zd}:
\ber
\frac{\mathrm{E}_{\mathrm{inel}}}{\mathrm{E}_0} &=& - 8\alpha^2 \int_0^\infty \frac{dQ^2}{Q^2} \int_{\nu^{\mathrm{inel}}_{\mathrm{thr}}}^\infty d\nu_\gamma
					\nonumber\\
& \times&	\hspace{-0.2cm}\left(	\frac{ \widetilde\gamma_1(\tilde{\tau},\tau_l) F_1(\nu_\gamma,Q^2)}{\nu_\gamma} +	\frac{ \widetilde\gamma_2(\tilde{\tau},\tau_l) F_2(\nu_\gamma,Q^2)}{4 M \tau_P} \right), \nonumber \\
\eer
with the photon energy $\nu_\gamma = \left( p \cdot q \right)/ M$, the virtuality $Q^2 = -q^2$ and kinematical notations:
\ber
 \tau_\mathrm{l} = \frac{Q^2}{4 m^2}, ~~~~~ \tau_\mathrm{P} = \frac{Q^2}{4 M^2}, ~~~~~  \tilde{\tau} = \frac{\nu_\gamma^2}{Q^2}.  \label{taus}
\eer
The photon-energy integration starts from the pion-nucleon inelastic threshold $ \nu^{\mathrm{inel}}_{\mathrm{thr}} $:
\ber
 \nu^{\mathrm{inel}}_{\mathrm{thr}} = m_\pi +  \frac{ m_\pi^2 + Q^2 }{ 2 M },
\eer
where $ m_{\pi} $ denotes the pion mass. The weighting functions $\tilde{\gamma}_1$ and $\tilde\gamma_2$ are given by \cite{Carlson:2011zd}
\ber
\tilde{\gamma}_1(\tau_1,\tau_2) &=& \frac{\sqrt{\tau_2} \gamma_1(\tau_2) - \sqrt{\tau_1} \gamma_1(\tau_1)}{\tau_2 - \tau_1},\nonumber\\
\tilde\gamma_2(\tau_1,\tau_2) &=& \frac{1}{\tau_2 - \tau_1}\left(\frac{ \gamma_2(\tau_1)}	{\sqrt{\tau_1}} - \frac{\gamma_2(\tau_2)}{\sqrt{\tau_2} } \right).
\eer

The contribution $ \mathrm{E}_{\mathrm{subt}}$ from the forward Compton scattering subtraction function $\mathrm{T}_1^\mathrm{subt} \left( 0,Q^2\right)$ is defined according to Refs. \cite{Pachucki:1996zza,Carlson:2011zd,Birse:2012eb,Tomalak:2015hva} as 
\ber
\frac{\mathrm{E}_{\mathrm{subt}}}{\mathrm{E}_0} &=& 4 \alpha M \int_0^\infty \frac{dQ^2}{Q^2}  \frac{\gamma_1\left( \tau_l \right)}{\sqrt{\tau_l}} \mathrm{T}_1^\mathrm{subt} \left( 0,Q^2\right),
\eer
with
\ber
\gamma_1(\tau) &=& \left(1-2\tau\right) \left(\left(1+\tau\right)^{1/2} - \tau^{1/2} \right) + \tau^{1/2},
\eer
and determined mainly by the value of the magnetic polarizability $\beta_M$ entering the low-energy expansion of $\mathrm{T}_1$ as 
\ber
\mathrm{T}_1^\mathrm{subt} \left( 0,Q^2\right) = \beta_M Q^2 + O \left( Q^4 \right).
\eer

The left part of the TPE effect from proton form factors is called the Born correction $ \mathrm{E}_{\mathrm{Born}}$ \cite{Pachucki:1996zza,Carlson:2011zd}:
\ber
\frac{\mathrm{E}_{\mathrm{Born}}}{\mathrm{E}_0}\hspace{-0.2cm} &=& \hspace{-0.2cm}4 \alpha^2 \int_0^\infty \frac{dQ^2}{Q^2}  \nonumber \\
&& \hspace{-0.2cm} \left(   - \frac{\gamma_1\left( \tau_l \right)}{\sqrt{\tau_l}} \left(  \mathrm{F}^2_\mathrm{D} - 1 \right) + \frac{16 M^2 m^2 \mathrm{G}_E' (0)}{ \left( M + m \right) Q}  \right. \nonumber \\
&+& \hspace{-0.2cm} \left. \frac{m^2 }{M^2 - m^2}  \left(  \left( \frac{ \gamma_1(\tau_P)}{\sqrt{\tau_P}} - \frac{ \gamma_1(\tau_l)}{\sqrt{\tau_l}} \right) \left(\mathrm{G}_M^2 -1\right) \right. \right. \nonumber \\
 &-& \hspace{-0.2cm} \left. \left.  \left( \frac{ \gamma_2(\tau_P)}{\sqrt{\tau_P}} - \frac{ \gamma_2(\tau_l)}{\sqrt{\tau_l}} \right) \frac{ \mathrm{G}_E^2 -1 + \tau_P \left(\mathrm{G}_M^2 -1\right) } { \tau_P (1+\tau_P) } \right)\right), \nonumber \\ \label{Born_Lamb}
\eer
with the Dirac ($ \mathrm{F}_\mathrm{D}$), Sachs electric ($ \mathrm{G}_\mathrm{E} $) and magnetic ($ \mathrm{G}_\mathrm{M} $) form factors. The kinematical factor $\gamma_2 $ is given by
\ber
\gamma_2(\tau) &=& \left(1+\tau \right)^{3/2} - \tau^{3/2} - \frac{3}{2}\tau^{1/2}.
\eer
In this term, we expand the electric form factor in terms of charge radius at low momentum transfer following evaluation of the Zemach correction in Ref. \cite{Tomalak:2017npu}, and the third Zemach moment contribution in Refs. \cite{Karshenboim:2014maa,Karshenboim:2015bwa}, and connect regions of large and small momentum transfer. We take an average of Refs. \cite{Antognini:1900ns,Bernauer:2013tpr} for a central value of the charge radius and estimate its uncertainty as half the difference between results in Refs. \cite{Antognini:1900ns,Bernauer:2013tpr} and provide the evaluation for the charge radius from the muonic hydrogen spectroscopy $r^{\mu \mathrm{H}}_\mathrm{E}$~\cite{Antognini:1900ns}. We evaluate the Born contribution exploiting form factors of Refs. \cite{Bernauer:2010wm,Bernauer:2013tpr} and take the unpolarized proton structure functions from the fit of Refs. \cite{Bosted:2007xd,Christy:2007ve,Donnachie:2004pi}.
\begin{table}[h] 
\begin{center}
\begin{tabular}{|c|c|c|c|}
\hline
 $\mathrm{E}^{e\mathrm{H}}_\mathrm{LS} $(1S) & $\mathrm{Hz}$   \\ \hline
Born, $ \mathrm{E}^{e\mathrm{H}}_{\mathrm{Born}}$ &   -44.1(9.6)   \\ 
Born, $ \mathrm{E}^{e\mathrm{H}}_{r^{\mu \mathrm{H}}_\mathrm{E}}$  &   -39.9(6.8)   \\ 
Subtraction, $\mathrm{E}^{e\mathrm{H}}_{\mathrm{subt}}$ &   18.5(4.4)   \\ 
Inelastic, $ \mathrm{E}^{e\mathrm{H}}_{\mathrm{inel}} $ &   $-$83.6(5.7)   \\ 
Total, $ \mathrm{E}^{e\mathrm{H}}_\mathrm{LS} =  \mathrm{E}^{e\mathrm{H}}_{\mathrm{Born}} + \mathrm{E}^{e\mathrm{H}}_{\mathrm{subt}} +  \mathrm{E}^{e\mathrm{H}}_{\mathrm{inel}} $ &   $-$109.2(12.0)   \\ \hline \hline
 $\mathrm{E}^{\mu\mathrm{H}}_\mathrm{LS} $(1S) & $\mathrm{\mu eV}$   \\ \hline
Born, $ \mathrm{E}^{\mu\mathrm{H}}_{\mathrm{Born}}$ &   -166.1(19.5)   \\ 
Born, $ \mathrm{E}^{\mu \mathrm{H}}_{r^{\mu \mathrm{H}}_\mathrm{E}}$  &   -148.9(12.8)   \\ 
Subtraction, $\mathrm{E}^{\mu\mathrm{H}}_{\mathrm{subt}}$ \cite{Tomalak:2015hva} &   18.5(10.0)   \\ 
Inelastic, $ \mathrm{E}^{\mu\mathrm{H}}_{\mathrm{inel}} $ \cite{Carlson:2011zd} &   $-$101.2(4.3)   \\ 
Total, $ \mathrm{E}^{\mu\mathrm{H}}_{\mathrm{LS}} =  \mathrm{E}^{\mu\mathrm{H}}_{\mathrm{Born}} + \mathrm{E}^{\mu\mathrm{H}}_{\mathrm{subt}} +  \mathrm{E}^{\mu\mathrm{H}}_{\mathrm{inel}} $ &   $-$248.8(22.3)    \\ \hline 
 %$\mathrm{E}^{\mu\mathrm{H}}_\mathrm{LS} $(1S) & $\mathrm{GHz}$   \\ \hline
%Born, $ \mathrm{E}^{\mu\mathrm{H}}_{\mathrm{Born}}$ &   -40.16(4.72)   \\ 
%Subtraction, $\mathrm{E}^{\mu\mathrm{H}}_{\mathrm{subt}}$ \cite{Tomalak:2015hva} &   4.49(2.42)   \\ 
%Inelastic, $ \mathrm{E}^{\mu\mathrm{H}}_{\mathrm{inel}} $ \cite{Carlson:2011zd} &   $-$24.6(1.0)   \\ 
%Total, $ \mathrm{E}^{\mu\mathrm{H}}_{\mathrm{2\gamma}} =  \mathrm{E}^{\mu\mathrm{H}}_{\mathrm{Born}} + \mathrm{E}^{\mu\mathrm{H}}_{\mathrm{subt}} +  \mathrm{E}^{\mu\mathrm{H}}_{\mathrm{inel}} $ &   $-$60.1(5.4)    \\ \hline 
\end{tabular}
\caption{Finite-size $\alpha^5$ TPE contributions to the Lamb shift of S energy levels in electronic and muonic hydrogen. } \label{results_LS}
\end{center}
\end{table}

We present results for TPE corrections to the Lamb shift (LS) of the ground state in electronic hydrogen in Tab. \ref{results_LS}. The Born TPE is around 1.3 times larger than the leading third Zemach momen correction \cite{Pachucki:1996zza}. The contribution from the subtraction function in electronic hydrogen is roughly two times smaller than the Born correction and larger than the estimate of Ref. \cite{Martynenko:2005rc}, where the smaller value of the proton magnetic polarizability $\beta_M = \left(1.9 \pm 0.5 \right) \times 10^{-4}~\mathrm{fm}^3 $, compared to the current p.d.g. quotation $\beta_M = \left(2.5 \pm 0.5 \right) \times 10^{-4}~\mathrm{fm}^3 $ \cite{Tanabashi:2018oca}, was used and the $Q^2$-dependence of the subtraction function was assumed but not remove well-justified by data or theory. The inelastic correction to the Lamb shift is almost twice larger than the Born contribution and 1.3 times larger than the result in the logarithmic approximation \cite{Pachucki:1996zza}. Our estimate is 1.1 times smaller than the calculation of Ref. \cite{Faustov:1999ga} and agrees with an update of Ref. \cite{Martynenko:2005rc} within uncertainties. In Ref. \cite{Faustov:1999ga}, the inelastic contribution was described by the Regge model. The model of structure functions as a sum of resonances with nonresonant background was used in Ref. \cite{Martynenko:2005rc}, while the result in Tab. \ref{results_LS} is based mainly on the fit of precise JLAB experimental data in the resonance region of Refs. \cite{Bosted:2007xd,Christy:2007ve}. Note that the inelastic two-photon effect in electronic hydrogen is in agreement within errors with the dispersive calculation of Ref. \cite{Rosenfelder:1999px} which is based mainly on the photoabsorption cross section data modified by empirical elastic form factors. The sum of inelastic and subtraction corrections is closer to the logarithmic approximation of Ref. \cite{Pineda:2004mx} than to the full heavy-baryon effective field theory calculation of Ref. \cite{Nevado:2007dd}. Moreover, we present results for the muonic hydrogen in Tab. \ref{results_LS}. The Born correction in muonic hydrogen is accidentally in a reasonable agreement with Ref. \cite{Pachucki:1996zza}, where the dipole parametrization of proton form factors was used, and slightly smaller than the previous estimate of Ref. \cite{Carlson:2011zd}, where we have combined proton state contributions in Ref. \cite{Carlson:2011zd} for comparison, due to our implementation of the expansion at low momentum transfer with the smaller charge radius value. Indeed, $ \mathrm{E}^{\mu\mathrm{H}}_{\mathrm{Born}}$ differs by $34.5~\mathrm{\mu eV}$ substituting the charge radius of Ref. \cite{Antognini:1900ns} versus Ref. \cite{Bernauer:2013tpr}.

Studying the isotope shift in light atoms, it is instructive to know also the two-photon effect due to the scattering on a single neutron \cite{Pachucki:2018yxe,Friar:2013rha}. We repeat the Lamb shift calculation without the subtraction of pure Coulomb part and leading charge radius ($\sim G'_E(0)$) contribution in Eq. (\ref{Born_Lamb}) in case of the neutron. Note that a special care has to be taken applying these results to nuclei, since we normalize to the energy $\mathrm{E}_0$ of Eq. (\ref{normalization_energy}) which changes going to the nucleus. We exploit form factors from Refs. \cite{Kubon:2001rj,Warren:2003ma,Kelly:2004hm,Punjabi:2015bba,Ye:2017gyb}, use the fit of Christy and Bosted \cite{Bosted:2007xd} for the unpolarized structure functions, and estimate the subtraction function following Ref. \cite{Tomalak:2015hva} with the neutron magnetic polarizability $ \beta_M = \left(3.7 \pm 1.2 \right)\times10^{-4}~\mathrm{fm}^3$ from p.d.g. \cite{Tanabashi:2018oca} and the Reggeon residue according to Refs. \cite{Pilkuhn:1973wq,Donnachie:2004pi,Gasser:2015dwa}. We present results in Tab. \ref{results_neutron}. 
\begin{table}[h] 
\begin{center}
\begin{tabular}{|c|c|c|c|}
\hline
 $\mathrm{E}^{e\mathrm{n}}_\mathrm{LS} $(1S) & $\mathrm{Hz}$   \\ \hline
Born, $ \mathrm{E}^{e\mathrm{n}}_{\mathrm{Born}}$ &  9.4(0.6)   \\ 
Subtraction, $\mathrm{E}^{e\mathrm{n}}_{\mathrm{subt}}$ &    28.8(10.2)    \\ 
Inelastic, $ \mathrm{E}^{e\mathrm{n}}_{\mathrm{inel}} $ &   $-$81.7(8.4)   \\ 
Total, $ \mathrm{E}^{e\mathrm{n}}_\mathrm{LS} =  \mathrm{E}^{e\mathrm{n}}_{\mathrm{Born}} + \mathrm{E}^{e\mathrm{n}}_{\mathrm{subt}} +  \mathrm{E}^{e\mathrm{n}}_{\mathrm{inel}} $ &   $-$43.5(13.2) \\ \hline 
\hline
 $\mathrm{E}^{\mu \mathrm{n}}_\mathrm{LS} $(1S) & $\mathrm{\mu eV}$   \\ \hline
Born, $ \mathrm{E}^{\mu \mathrm{n}}_{\mathrm{Born}}$ &   10.5(2.9)   \\ 
Subtraction, $\mathrm{E}^{\mu \mathrm{n}}_{\mathrm{subt}}$ &  34.6(15.6)   \\ 
Inelastic, $ \mathrm{E}^{\mu \mathrm{n}}_{\mathrm{inel}} $ &   $-$102.9(8.6)   \\ 
Total, $ \mathrm{E}^{\mu \mathrm{n}}_\mathrm{LS} =  \mathrm{E}^{\mu \mathrm{n}}_{\mathrm{Born}} + \mathrm{E}^{\mu \mathrm{n}}_{\mathrm{subt}} +  \mathrm{E}^{\mu \mathrm{n}}_{\mathrm{inel}} $ &   $-$57.8(18.1)  \\ \hline 
\end{tabular}
\caption{Finite-size $\alpha^5$ TPE contributions to the Lamb shift from electromagnetic interaction with neutron $\mathrm{E}_\mathrm{LS}^{e\mathrm{n}},~\mathrm{E}_\mathrm{LS}^{\mu\mathrm{n}}$.} \label{results_neutron}
\end{center}
\end{table}
The Born correction in $e n$ and $\mu n$ systems has a different sign compared to $e p$ and $\mu p$. For a neutron with zero charge, the elastic Friar term is relatively small compared to the positively charged proton, and the main contribution comes from the neutron magnetic form factor resulting in a positive sign and relatively small uncertainty. We obtain the central value averaging over the form factor parametrizations and estimate the uncertainty as a difference between the largest and smallest results. As in Ref. \cite{Pachucki:2018yxe}, the inelastic corrections for proton and neutron coincide within errors. We double the uncertainty for the inelastic contribution in case of the neutron compared to the proton. Note that the resulting two-photon exchange effect in $\mu \mathrm{H}$ is roughly four times larger than in $\mu n$ system: $\mathrm{E}^{\mu \mathrm{H}}_\mathrm{LS} \approx 4 \mathrm{E}^{\mu \mathrm{n}}_\mathrm{LS} $, as it has been estimated in Refs. \cite{Krauth:2015nja,Pachucki:2018yxe}. The main uncertainty in the two-photon correction is due to the pure knowledge of the forward Compton scattering subtraction function. However, it can be improved exploiting the chiral perturbation theory predictions \cite{Birse:2012eb,Alarcon:2013cba,Peset:2014jxa}, constraints at high energy \cite{Collins:1978hi,Hill:2016bjv} as well as the phenomenological studies of the difference between the subtraction function for protons and neutrons \cite{Gasser:1974wd,Gasser:2015dwa}, and by improved extraction of the neutron magnetic polarizability \cite{Lundin:2002jy,Kossert:2002ws,Myers:2014ace,Annand:2013ace}.

For the hyperfine splitting correction $\mathrm{E}_{\mathrm{HFS}}$, we use definitions of Refs. \cite{Eides:2000xc,Tomalak:2017lxo,Tomalak:2017npu,Tomalak:2017owk}. The result is given by a sum of the Zemach $\mathrm{E}_{\mathrm{Z}}$, the recoil $\mathrm{E}_{\mathrm{R}}$ and the polarizabillity $\mathrm{E}_{\mathrm{pol}}$ terms:
\ber
\mathrm{E}_{\mathrm{HFS}} = \mathrm{E}_{\mathrm{Z}} + \mathrm{E}_{\mathrm{R}} + \mathrm{E}_{\mathrm{pol}},
\eer
where the contributions relative to the leading Fermi splitting $\mathrm{E}_\mathrm{F}$:
\ber
\mathrm{E}_\mathrm{F} = \frac{8 \pi \alpha}{3}   \mu_P \frac{|\psi_{\mathrm{nS}}\left(0\right)|^2}{M m},
\eer
with the proton magnetic moment $\mu_P$ are given by
\ber \label{zemach_correction}
\frac{\mathrm{E}_{\mathrm{Z}}}{\mathrm{E}_\mathrm{F}}& = &\frac{8 \alpha m_\mathrm{r}}{\pi }  \int \limits^{\infty}_{0} \frac{\mathrm{d} Q}{Q^2} \left( \frac{\mathrm{G}_\mathrm{E}\left(Q^2\right) \mathrm{G}_\mathrm{M}\left(Q^2\right)  }{\mathrm{\mu}_\mathrm{P}}  - 1 \right), \\
\frac{\mathrm{E}_{\mathrm{R}}}{\mathrm{E}_\mathrm{F}} &=&\frac{\alpha}{\pi} \int \limits^{\infty}_{0} \frac{\mathrm{d} Q^2}{Q^2}  \frac{ \left( 2 + \rho\left(\tau_\mathrm{l}\right) \rho\left(\tau_\mathrm{P} \right)  \right) \mathrm{F}_\mathrm{D}\left(Q^2\right)  }{ \sqrt{\tau_\mathrm{P}} \sqrt{1+\tau_\mathrm{l}} + \sqrt{\tau_\mathrm{l}} \sqrt{1+\tau_\mathrm{P}} } \frac{ \mathrm{G}_\mathrm{M}\left(Q^2\right)  }{\mathrm{\mu}_\mathrm{P}} \nonumber \\ 
&+&  \frac{3 \alpha}{\pi} \int \limits^{\infty}_{0} \frac{\mathrm{d} Q^2}{Q^2}  \frac{ \rho\left(\tau_\mathrm{l}\right) \rho\left(\tau_\mathrm{P} \right) \mathrm{F}_\mathrm{P}\left(Q^2\right)  }{ \sqrt{\tau_\mathrm{P}} \sqrt{1+\tau_\mathrm{l}} + \sqrt{\tau_\mathrm{l}} \sqrt{1+\tau_\mathrm{P}} } \frac{ \mathrm{G}_\mathrm{M}\left(Q^2\right)  }{\mathrm{\mu}_\mathrm{P}}  \nonumber \\
&-& \frac{\alpha}{\pi} \int \limits^{\infty}_{0} \frac{\mathrm{d} Q}{Q}\left(\frac{m}{M} \frac{\rho (\tau_l) \left( \rho (\tau_l) - 4 \right)\mathrm{F}^2_\mathrm{P} \left(Q^2\right)}{\mathrm{\mu}_\mathrm{P}} - \frac{8 m_\mathrm{r}}{Q} \right) \nonumber \\
&-& \Delta_Z,  \label{recoil_correction} \\
\frac{\mathrm{E}_{\mathrm{pol}}}{\mathrm{E}_\mathrm{F}}\hspace{-0.05cm} &=& \hspace{-0.05cm}\frac{2 \alpha}{\pi \mathrm{\mu}_\mathrm{P}} \int \limits^{\infty}_{0} \frac{\mathrm{d} Q^2}{Q^2} \int \limits^{\infty}_{\nu^{\mathrm{inel}}_{\mathrm{thr}}} \frac{\mathrm{d} \nu_\gamma}{\nu_\gamma}  \frac{\left( 2 + \rho\left(\tau_\mathrm{l}\right) \rho\left(\tilde{\tau}\right) \right)  g_1 \left(\nu_\gamma, Q^2 \right)}{  \sqrt{\tilde{\tau}} \sqrt{1+\tau_\mathrm{l}} + \sqrt{\tau_\mathrm{l}} \sqrt{1+\tilde{\tau}}   } \nonumber \\
 &-& \frac{6 \alpha}{\pi \mathrm{\mu}_\mathrm{P}} \int \limits^{\infty}_{0} \frac{\mathrm{d} Q^2}{Q^2} \int \limits^{\infty}_{\nu^{\mathrm{inel}}_{\mathrm{thr}}} \frac{\mathrm{d} \nu_\gamma}{ \nu_\gamma} \frac{1}{\tilde{\tau}}  \frac{ \rho\left(\tau_\mathrm{l}\right) \rho\left(\tilde{\tau} \right)  g_2 \left(\nu_\gamma, Q^2 \right) }{  \sqrt{\tilde{\tau}} \sqrt{1+\tau_\mathrm{l}} + \sqrt{\tau_\mathrm{l}} \sqrt{1+\tilde{\tau}}   } \nonumber \\
&+&  \frac{\alpha}{\pi} \int \limits^{\infty}_{0} \frac{\mathrm{d} Q}{Q}\frac{m}{M}  \frac{\rho (\tau_l) \left( \rho (\tau_l) - 4 \right) \mathrm{F}^2_P \left(Q^2\right)}{\mu_P} , \label{polar_correction}
\eer
with $\rho(\tau) = \tau - \sqrt{\tau ( 1 + \tau )}$, $ \mathrm{F}_\mathrm{P}$ is the Pauli form factor, $g_1 \left(\nu_\gamma, Q^2 \right)$ and $g_2 \left(\nu_\gamma, Q^2 \right)$ are the spin-dependent inelastic proton structure functions. For the electronic hydrogen, the Zemach correction $ \mathrm{E}_{\mathrm{Z}}$ is obtained by scaling with the reduced mass in hydrogen to the muonic hydrogen from the averaged over electric and magnetic radii result of Ref. \cite{Tomalak:2017npu}. The recoil $ \mathrm{E}_{\mathrm{R}} $ and polarizability $ \mathrm{E}_{\mathrm{pol}} $ contributions are evaluated following the same steps as in Ref. \cite{Tomalak:2017npu} for $\mu \mathrm{H}$. The proton spin structure functions parametrization is based on Refs. \cite{Prok:2008ev,Kuhn:2008sy,griffioen,Sato:2016tuz,Fersch:2017qrq}. 

 In Tab. \ref{results_HFS}, we provide the hyperfine-splitting TPE contributions as well as extractions from the experimental data exploiting radiative corrections of Refs. \cite{Bodwin:1987mj,Ivanov:1996ew,Karshenboim:1996ew,Eides:2000xc,Martynenko:2004bt,Peset:2016wjq,Jegerlehner:2011mw,Jegerlehner:2017pr,Dorokhov:2017gst,Faustov:2017hfo}. The experimental value of the hyperfine splitting in muonic hydrogen is taken from Ref. \cite{Antognini:1900ns} and in electronic hydrogen from Refs. \cite{Hellwig:1970,Zitzewitz:1970,Essen:1971x,Morris:1971,Essen:1973,Reinhard:1974,Vanier:1976,Petit:1980,Cheng:1980,Karshenboim:2005iy,Horbatsch:2016xx}.
\begin{table}[h] 
\begin{center}
\begin{tabular}{|c|c|c|c|}
\hline
 $\mathrm{E}^{e\mathrm{H}}_\mathrm{HFS} $(1S) & $\mathrm{kHz}$   \\ \hline
Zemach, $ \mathrm{E}^{e\mathrm{H}}_{\mathrm{Z}} $ &   $-$56.54(70) \\  
Recoil, $ \mathrm{E}^{e\mathrm{H}}_{\mathrm{R}} $  & 7.56(7)  \\ 
Polarizability, $ \mathrm{E}^{e\mathrm{H}}_{\mathrm{pol}}  $  &  2.71(0.72)    \\ 
HFS, $ \mathrm{E}^{e\mathrm{H}}_{\mathrm{HFS}} = \mathrm{E}^{e\mathrm{H}}_{\mathrm{Z}}+\mathrm{E}^{e\mathrm{H}}_{\mathrm{R}} +  \mathrm{E}^{e\mathrm{H}}_{\mathrm{pol}} $ &   $-$46.27(1.09)   \\ 
$ \mathrm{E}^{e\mathrm{H}}_{\mathrm{HFS}} $ from 1S HFS in eH &   $-$46.14((1),~(34))    \\ \hline \hline
 $\mathrm{E}^{\mu\mathrm{H}}_\mathrm{HFS} $(1S) & $\mathrm{\mu eV}$   \\  \hline
Zemach, $ \mathrm{E}^{\mu\mathrm{H}}_{\mathrm{Z}} $ \cite{Tomalak:2017npu}  &   $-$1352(17) \\  
Recoil, $ \mathrm{E}^{\mu\mathrm{H}}_{\mathrm{R}} $  \cite{Tomalak:2017npu}  & 154(1)  \\ 
Polarizability, $ \mathrm{E}^{\mu\mathrm{H}}_{\mathrm{pol}} $  \cite{Tomalak:2017npu}   &  66(16)    \\ 
$ \mathrm{E}^{\mu\mathrm{H}}_{\mathrm{HFS}} =   \mathrm{E}^{\mu\mathrm{H}}_{\mathrm{Z}}+\mathrm{E}^{\mu\mathrm{H}}_{\mathrm{R}} +  \mathrm{E}^{\mu\mathrm{H}}_{\mathrm{pol}}$ \cite{Tomalak:2017npu} &   $-$1131(24)    \\  
$ \mathrm{E}^{\mu\mathrm{H}}_{\mathrm{HFS}} $ from 2S HFS in $\mu\mathrm{H}$ &   $-$1162((41),~(42))    \\  
$ \mathrm{E}^{\mu\mathrm{H}}_{\mathrm{HFS}} $ from 1S HFS in $e\mathrm{H}$ &   $-$1127.6((3.6),~(9.0))    \\ \hline 
\end{tabular}
\caption{Finite-size $\alpha^5$ TPE contributions to the hyperfine splitting of S energy levels in hydrogen and muonic hydrogen. In experimental extractions, the first uncertainty is the error of radiative corrections and measurement, and the second one contains a possible $ \alpha \mathrm{E}_{\mathrm{HFS}}$ error from higher orders. } \label{results_HFS}
\end{center}
\end{table} 
All corrections to the hyperfine splitting in electronic hydrogen are three orders of magnitude above the Lamb shift contributions. As well as in muonic hydrogen \cite{Tomalak:2017npu}, they slightly differ to the previous estimates of Ref. \cite{Carlson:2008ke} due to the inclusion of the recent form factor measurements \cite{Bernauer:2010wm,Bernauer:2013tpr}. Theoretical estimates of the hyperfine-splitting correction are within errors of the phenomenological extraction from measurements.

Additionally, we provide an update of Ref. \cite{Tomalak:2017lxo} for the absolute value of the hyperfine-splitting energy $ \mathrm{E}^{\mu \mathrm{H}}_{\mathrm{HFS}}$ in muonic hydrogen removing axial-vector mesons \cite{Dorokhov:2017nzk} from the analysis and accounting for the vacuum polarization graphs with elastic and inelastic proton structure in higher-order radiative corrections:
 \ber
 \mathrm{E}^{\mu \mathrm{H}}_{\mathrm{HFS}} \left( \mathrm{1S} \right) &=& 182.625 \pm 0.012~\mathrm{meV}, \\
 \mathrm{E}^{\mu \mathrm{H}}_{\mathrm{HFS}} \left( \mathrm{2S} \right)  &=& 22.8132 \pm 0.0015~\mathrm{meV}.
 \eer
An improved calculation of two-photon diagrams with QED corrections on fermion lines, graphs with three exchanged photons  \cite{Kalinowski:2018vjm} as well as evaluation of the two-photon contributions in non-forward kinematics can reduce the uncertainty further.
 
We presented the current knowledge of the TPE correction to S energy levels. The Lamb shift results can be useful in future extractions of the isotope shift, while the contributions to the hyperfine splitting can help to tune and analyze forthcoming 1S HFS measurements in $\mu\mathrm{H}$~\cite{Pohl:2016tqq,Dupays:2003zz,Adamczak:2016pdb,Ma:2016etb}.

We acknowledge Krzysztof Pachucki for the advice given during the manuscript preparation and Marcin Kalinowski for useful discussion. This work was supported in part by a NIST precision measurement grant and by the U. S. Department of Energy, Office of Science, Office of High Energy Physics, under Award No. DE-SC0019095. This work was supported by the Deutsche Forschungsgemeinschaft (DFG) through Collaborative Research Center ``The Low-Energy Frontier of the Standard Model'' (SFB 1044). The author would like to acknowledge the Mainz Institute for Theoretical Physics (MITP) for its hospitality and support.

\end{document}